\documentclass[twocolumn, twocolappendix,times, preprint]{aastex6}

\usepackage{amsmath, bm}

\begin{document}


\title{The magnitude of viscous dissipation in strongly stratified two-dimensional convection}

%% Use \author, \affil, plus the \and command to format author and affiliation 
%% information.  If done correctly the peer review system will be able to
%% automatically put the author and affiliation information from the manuscript
%% and save the corresponding author the trouble of entering it by hand.
%%
%% The \affil should be used to document primary affiliations and the
%% \altaffil should be used for secondary affiliations, titles, or email.

%% Authors with the same affiliation can be grouped in a single
\author{Laura K. Currie and Matthew K. Browning}
\affil{Department of Physics and Astronomy, University of Exeter, Stocker Road, EX4 4QL Exeter, UK; lcurrie@astro.ex.ac.uk}
%\altaffiltext{lcurrie@astro.ex.ac.uk}
%\altaffiltext{Department of Applied Mathematics, University of Leeds, Leeds, UK, LS2 9JT}

%% Use the \and command so offset the last author.
%%\and

%\author{Jeff Lewandowski\altaffilmark{5}}
%\affil{IOP Publishing, Washington, DC 20005}


%\altaffiltext{1}{AAS Journals Data Scientist}
%\altaffiltext{2}{greg.schwarz@aas.org}
%\altaffiltext{3}{AAS Journals Associate Editor-in-Chief}
%\altaffiltext{4}{AAS Director of Publishing}
%\altaffiltext{5}{IOP Senior Publisher for the AAS Journals}

\begin{abstract}
Convection in astrophysical systems must be maintained against dissipation. Although the effects of dissipation are often assumed to be negligible, theory suggests that in strongly stratified convecting fluids, the dissipative heating rate can exceed the luminosity carried by convection. Here we explore this possibility using a series of numerical simulations. We consider two-dimensional numerical models of hydrodynamic convection in a Cartesian layer under the anelastic approximation and demonstrate that the dissipative heating rate can indeed exceed the imposed luminosity. We establish a theoretical expression for the ratio of the dissipative heating rate to the luminosity emerging at the upper boundary, in terms only of the depth of the layer and the thermal scale height. In particular, we show that this ratio is independent of the diffusivities and confirm this with a series of numerical simulations. Our results suggest that dissipative heating may significantly alter the internal dynamics of stars and planets.  
\end{abstract}




\section{Introduction}
Convection occurs in the interiors of many astrophysical bodies and must be sustained against viscous and ohmic dissipation. This dissipation is often neglected in astrophysical models, e.g., in standard stellar 1D evolution codes \citep[e.g.,][]{ChabrierBaraffe1997,Paxtonetal2011} though its effects have lately been considered in a few specific contexts \citep[e.g.,][]{BatyginStevenson2010,Browningetal2016}.

Astrophysical convection often occurs over many scale heights. While for incompressible fluids the contribution of dissipative heating to the internal energy budget is negligible \citep{Kundu1990}, \citet{Hewittetal1975} (hereafter HMW) showed that in strongly stratified systems, it is theoretically possible for the rate of dissipative heating to exceed the luminosity. This was supported numerically by \citet{JarvisMcKenzie1980} for the case of a compressible liquid with infinite Prandtl number, $Pr$, (the ratio of viscous and thermal diffusivities), appropriate for models of the Earth's interior.

In this study we aim to establish the magnitude of dissipation for conditions more akin to those encountered in stellar interiors. Specifically, we consider dissipation in a stratified gas at finite Pr, and examine how the total heating changes as system parameters are varied. To begin, we briefly review some relevant thermodynamic considerations that underpin our work.
 
%%%%%%
 
\subsection{Thermodynamic constraints on dissipative heating}\label{Hewitt}
For a volume $V$ of convecting fluid enclosed by a surface $S$ with associated magnetic field $\mathbf{B}$, in which the normal component of the fluid velocity $\mathbf{u}$ vanishes on the surface, and either all components of $\mathbf{u}$, or the tangential stress, also vanish on the surface, local conservation of energy gives that the rate of change of total energy is equal to the sum of the net inward flux of energy and the rate of internal heat generation (e.g., by radioactivity or nuclear reactions). This implies 
\begin{align}\label{consofE}\frac{\partial}{\partial{t}}\left(\rho{e}+\frac{1}{2}\rho{u}^2\right.&\left.+\frac{B^2}{2\mu_0}-\rho\Psi\right)=-\nabla\cdot\left(\rho\left(e+\frac{1}{2}u^2-\Psi\right)\mathbf{u}\right.\nonumber\\&\left.+\frac{(\mathbf{E}\times\mathbf{B})}{\mu_0}+P\mathbf{u}-\bm\tau\cdot\mathbf{u}-k\nabla{T}\right)+H\end{align}
where $\rho$ is the fluid density, $e$ is the internal energy of the fluid, $\Psi$ is the gravitational potential that satisfies $\mathbf{g}=\nabla\Psi$, $P$ is the pressure, $\tau_{ij}$ is the contribution to the total stress tensor from irreversible processes, $k$ is the thermal conductivity, $T$ is the temperature, $H$ is the rate of internal heat generation, and $\frac{\mathbf{E}\times\mathbf B}{\mu_0}$ is the Poynting flux ($\mathbf{E}$ is the electric field and $\mu_0$ is the permeability of free space). 
Integrating (\ref{consofE}) over $V$ gives the global relation
 \begin{equation}\label{Fbal}
\int_Sk\frac{\partial{T}}{\partial{x_i}}\,dS_i+\int_VH\,dV=0,
 \end{equation}
assuming both a steady state and that the electric current, $\mathbf{j}$, vanishes everywhere outside $V$. Equation (\ref{Fbal}) implies that the net flux out of $V$ is equal to the total rate of internal heating. Viscous and ohmic heating do not contribute to the overall heat flux: dissipative heating terms do not appear in equation (\ref{Fbal}).

To examine dissipative heating, we consider the internal energy equation:
\begin{equation}\label{internal}
\rho\left(\frac{\partial{e}}{\partial{t}}+(\mathbf{u}\cdot\nabla)e\right)=\nabla(k\nabla{T})-P(\nabla\cdot\mathbf{u})+\tau_{ij}\frac{\partial{u_i}}{\partial{x_j}}+\frac{j^2}{\sigma}+H
\end{equation}
where $\sigma$ is the conductivity of the fluid.
Integrating over $V$, and assuming a steady state, (\ref{internal}) becomes
\begin{equation}\label{Phibal}
\int_V(\mathbf{u}\cdot\nabla)P\,dV+\Phi =0.
\end{equation}
Here 
\begin{equation}\label{Phi}
\Phi=\int_V\tau_{ij}\frac{\partial{u_i}}{\partial{x_j}}+\frac{j^2}{\sigma}\,dV
\end{equation}
is the total dissipative heating rate including viscous and ohmic heating terms.
Equation (\ref{Phibal}) implies that the global rate of dissipative heating is cancelled by the work done against the pressure gradient. Equation (\ref{Phibal}) is only equivalent to HMW's equation (22) when considering an ideal gas (so that $\alpha{T}=1$, where $\alpha$ is the coefficient of thermal expansion); however, in arriving at (\ref{Phibal}), we made no assumption about the fluid being a gas. \citet{AlboussiereRicard2013,AlboussiereRicard2014} note that this inconsistency arises because HMW assume $c_p$ to be constant in their derivation, which is not valid when $\alpha T\neq1$.

Alternatively, from the first law of thermodynamics, we have
\begin{equation}
Tds=de-\frac{P}{\rho^2}d\rho
\end{equation}
where $s$ is the specific entropy, so (\ref{Phibal}) can also be written as 
\begin{equation}\label{Phi2}
\Phi=\int_V\rho{T}(\mathbf{u}\cdot\nabla)s\,dV=-\int_V\rho{s}(\mathbf{u}\cdot\nabla)T\,dV
\end{equation}
where we have invoked mass continuity in a steady state ($\nabla\cdot(\rho\mathbf{u})=0$).
Hence the global dissipation rate can also be thought of as being balanced by the work done against buoyancy \citep{JonesKuzanyan2009}.

HMW used the entropy equation to derive an upper bound for the dissipative heating rate in a steadily convecting fluid that is valid for any equation of state or stress-strain relationship. 
For the case of convection in a plane layer, that upper bound is 
\begin{equation}\label{bound}
\frac{\Phi}{L_u}<\frac{T_{max}-T_u}{T_u}
\end{equation}
where $L_u$ is the luminosity at the upper boundary, $T_{max}$ is the maximum temperature and $T_u$ is the temperature on the upper boundary. 

One consequence of this bound is that, for large enough thermal gradients, the dissipative heating rate may exceed the heat flux through the layer; this is perhaps counter-intuitive, but is thermodynamically permitted, essentially because the dissipative heating remains in the system's internal energy \citep[see e.g.,][]{Backus1975}.   

The above considerations should hold for both ohmic and viscous dissipation. However, HMW further considered the simple case of viscous heating in a liquid (neglecting magnetism) and showed that the viscous dissipation rate is not only bounded by (\ref{bound}) but that
\begin{equation}\label{Hewittliq}
E\equiv\frac{\Phi}{L_u}=\frac{d}{H_T}\left(1-\frac{\mu}{2}\right)   
\end{equation}
where $d$ is the height of the convective layer, $H_T$ is the (constant) thermal scale height and $0\leq\mu\leq1$ is the fraction of internal heat generation.
Interestingly, the theoretical expression (\ref{Hewittliq}) is dependent only on the ratio of the layer depth to the thermal scale height and the fraction of internal heat generation.

As expected, (\ref{Hewittliq}) implies that the dissipative heating rate is negligible when compared with the heat flux in cases where the Boussinesq approximation is valid (i.e., when the scale heights of the system are large compared to the depth of the motion).
But it follows from (\ref{Hewittliq}) that $\Phi$ is significant compared to $L_u$ if $d$ is comparable to $H_T$, i.e., if the system has significant thermal stratification. Stellar convection often lies in this regime, so it is not clear that dissipative heating can be ignored.

This paper explores these theoretical predictions using simulations of stratified convection under conditions akin to those encountered in stellar interiors.   Previous numerical simulations conducted by HMW considered only 2D Boussinesq convection and neglected inertial forces (infinite $Pr$ approximation); later work by \citet{JarvisMcKenzie1980} within the so-called anelastic liquid approximation considered stronger stratifications but likewise assumed a liquid at infinite $Pr$.  We extend these by considering an ideal gas (so that $\alpha{T}=1$) at finite $Pr$, so inertial effects are important and compressibility is not negligible. 

In section \ref{model}, we describe the model setup before presenting results from numerical simulations. In section \ref{discussion} we offer a discussion of the most significant results that emerge before providing conclusions.
%%%%%%%%

\section{Simulations of dissipative convection}\label{model}
\subsection{Model setup}\label{modelsec}
We consider a layer of convecting fluid lying between impermeable boundaries at $z=0$ and $z=d$. We assume thermodynamic quantities to be comprised of a background, time-independent, reference state and perturbations to this reference state. The reference state is taken to be a polytropic, ideal gas with polytropic index $m$ given by 
\begin{equation}\label{refstate}
\bar{T}=T_0(1-\beta z),\,\bar\rho=\rho_0(1-\beta z)^m,\,\bar{p}=\mathcal{R}\rho_0T_0(1-\beta z)^{m+1},
\end{equation}
where $\beta=\frac{g}{c_{p,0}T_0}$. Here, $g$ is the acceleration due to gravity, $c_p$ is the specific heat capacity at constant pressure, $\mathcal{R}$ is the ideal gas constant and a subscript $0$ represents the value of that quantity on the bottom boundary. $\beta$ is equivalent to the inverse temperature scale height and so is a measure of the stratification of the layer, although we shall use the more conventional 
\begin{equation}
N_{\rho}=-m\ln(1-\beta d)
\end{equation}
to quantify the stratification, with $N_{\rho}$ the number of density scale heights across the layer. We assume a polytropic, monatomic, adiabatic, ideal gas, therefore $m=1.5$. Here we consider only the hydrodynamic problem; i.e., all dissipation is viscous.

We use anelastic equations under the Lantz-Braginsky-Roberts (LBR) approximation \citep{Lantz1992,BraginskyRoberts1995}; these are valid when the reference state is nearly adiabatic and when the flows are subsonic \citep{OguraPhillips1962,Gough1969,LantzFan1999}, as they are here.

The governing equations are then
\begin{align}\frac{\partial\mathbf u}{\partial{t}}&+(\mathbf{u}\cdot\nabla)\mathbf{u}=-\nabla\tilde{p}+\frac{gs}{c_p}\hat{\mathbf{e_z}}\nonumber\\&+\nu\left[\frac{1}{\bar\rho}\frac{\partial}{\partial{x_j}}\left(\bar\rho\left(\frac{\partial{u_i}}{\partial{x_j}}+\frac{\partial{u_j}}{\partial{x_i}}\right)\right)-\frac{2}{3\bar\rho}\frac{\partial}{\partial{x_i}}\left(\bar\rho\frac{\partial{u_j}}{\partial{x_j}}\right)\right]\end{align} 
\begin{equation}
\nabla\cdot(\bar\rho\mathbf u)=0
\end{equation}
\begin{equation}\label{energyeq}
\bar\rho\bar{T}\left(\frac{\partial{s}}{\partial{t}}+(\mathbf{u}\cdot\nabla)s\right)=\nabla\cdot(\kappa\bar\rho\bar{T}\nabla{s})+\tau_{ij}\frac{\partial{u_i}}{\partial{x_j}}+H,
\end{equation}
where $\mathbf{u}$ is the fluid velocity, $\tilde{p}=\frac{p}{\bar\rho}$ is a modified pressure and $\nu$ is the kinematic viscosity. The specific entropy, $s$, is related to pressure and density by
\begin{equation}
s=c_v\ln{p}-c_p\ln\rho.
\end{equation}
We assume the perturbation of the thermodynamic quantities to be small compared with their reference state value. Therefore the entropy is obtained from
\begin{equation}
s=c_v\frac{p}{\bar{p}}-c_p\frac{\rho}{\bar\rho}
\end{equation}
and the linearised equation of state is
\begin{equation}
\frac{p}{\bar{p}}=\frac{T}{\bar{T}}+\frac{\rho}{\bar\rho}.
\end{equation}
In (\ref{energyeq}) $\kappa$ is the thermal diffusivity and \begin{equation}\label{tau}
\tau_{ij}=\nu\bar\rho\left(\frac{\partial{u_i}}{\partial{x_j}}+\frac{\partial{u_j}}{\partial{x_i}}-\frac{2}{3}\delta_{ij}\nabla\cdot\mathbf{u}\right)
\end{equation}
is the viscous stress tensor ($\delta_{ij}$ is the Kronecker delta).
Here, we only consider cases with $H=0$ (i.e., no internal heat generation), and instead impose a flux ($F$) at the bottom boundary.
Note the LBR approximation diffuses entropy (not temperature); see \cite{Lecoanetetal2014} for a discussion of the differences.
We assume a constant $\nu$ and $\kappa$. 

We solve these equations using the Dedalus pseudo-spectral code \citep{dedalus} with fixed flux on the lower boundary and fixed entropy on the upper boundary. 
We assume these boundaries to be impermeable and stress-free. We employ a sin/cosine decomposition in the horizontal, ensuring there is no lateral heat flux.
We employ the semi-implicit Crank-Nicolson Adams-Bashforth numerical scheme and typically use 192 grid points in each direction with dealiasing (so that 128 modes are used). In some cases, 384 (256) grid points (modes) were used to ensure adequate resolution of the solutions.
For simplicity, and to compare our results with those of HMW, we consider 2D solutions so that $\mathbf{u}=(u,0,w)$ and $\frac{\partial}{\partial{y}}\equiv0$. This also allows us to reach higher supercriticalities and $N_{\rho}$ with relative ease.

As we neglect magnetism, the total dissipation rate, $\Phi$, is given by (\ref{Phi}) with $\mathbf j=0$ and $\tau_{ij}$ as given by (\ref{tau}).

An appropriate non-dimensionalisation of the system allows the parameter space to be collapsed such that the dimensionless solutions (in particular $E$) are fully specified by $m$, $N_{\rho}$, $Pr$, together with $\hat{F_0}= \frac{Fd}{\kappa{c_{p,0}}\rho_0T_0}$ (a dimensionless measure of the flux applied at the lower boundary) and a flux-based Rayleigh number \citep[e.g.,][]{Duarteetal2016}      
\begin{equation}\label{Ra}
Ra=\frac{gd^4F_{u}}{\nu\kappa^2\rho_0c_{p,0}T_0}.
\end{equation}
The parameters used in our simulations are given in Table \ref{table1}.  

In a steady state, an expression for the luminosity $L$ at each depth $z=z'$ can be obtained by integrating the internal energy equation (\ref{energyeq}) over the volume contained between the bottom of the layer and the depth $z=z'$:
\begin{align}L=&FA=\int_{V_{z'}}\nabla\cdot(\bar\rho\bar{T}s\mathbf{u})\,dV+\int_{V_{z'}}-\nabla\cdot(\kappa\bar\rho\bar{T}\nabla{s})\,dV\nonumber\\&+\int_{V_{z'}}-s\bar\rho(\mathbf{u}\cdot\nabla)\bar{T}\,dV+\int_{V_{z'}}-\tau_{ij}\frac{\partial{u_i}}{\partial{x_j}}\,dV,\label{Feqpre}\end{align}
where $A$ is the surface area.
The divergence theorem allows the first two integrals to be transformed into surface integrals giving
\begin{align}L=&FA=\underbrace{\int_{S_{z'}}\bar\rho\bar{T}sw\,dS}_\text{$L_{conv}=AF_{conv}$}+\underbrace{\int_{S_{z'}}-\kappa\bar\rho\bar{T}\frac{\partial{s}}{\partial{z}}\,dS}_\text{$L_{cond}=AF_{cond}$}\nonumber\\&+\underbrace{\int_{V_{z'}}-s\bar\rho(\mathbf{u}\cdot\nabla)\bar{T}\,dV}_\text{$L_{buoy}=A\int_0^{z'}Q_{buoy}\,dz$}+\underbrace{\int_{V_{z'}}-\tau_{ij}\frac{\partial{u_i}}{\partial{x_j}}\,dV}_\text{$L_{diss}=A\int_0^{z'}Q_{diss}\,dz$},\label{Feq}\end{align}
where the surface integrals are over the surface at height $z=z'$.
The first and second terms define the horizontally-averaged heat fluxes associated with convection ($F_{conv}$) and conduction ($F_{cond}$) respectively, along with associated luminosities.
The third and fourth terms define additional sources of heating and cooling ($Q_{diss}$ and $Q_{buoy}$) associated with viscous dissipation and with work done against the background stratification, respectively. These two terms must cancel in a global sense i.e., when integrating from $z=0$ to $z=d$, but they do not necessarily cancel at each layer depth. 

An alternative view of the heat transport may be derived by considering the total energy equation (\ref{consofE}), which includes both internal and mechanical energy. 
In a steady state (with entropy diffusion), the local balance gives
\begin{equation}
\nabla\cdot\left(\bar\rho\left(e+\frac{1}{2}u^2-\Psi\right)\mathbf{u}+p\mathbf{u}-\bm\tau\cdot\mathbf{u}-\kappa\bar\rho{T}\nabla{s}\right)=H
\end{equation}
which when integrated over the volume for an ideal gas gives \citep[see e.g.,][]{Vialletetal2013}
\begin{align}L=&FA=\underbrace{\int_{S_{z'}}\bar\rho{c_p}wT'\,dS}_\text{$L_e=AF_{e}$}+\underbrace{\int_{S_{z'}}-\kappa\bar\rho\bar{T}\frac{\partial{s}}{\partial{z}}\,dS}_\text{$L_{cond}=AF_{cond}$}\nonumber\\&+\underbrace{\int_{S_{z'}}\frac{1}{2}\bar\rho|u^2|w\,dS}_\text{$L_{KE}=AF_{KE}$}+\underbrace{\int_{S_{z'}}-(\tau_{ij}{u_i})\cdot{\mathbf{\hat{e}_z}}\,dS}_\text{$L_{visc}=AF_{visc}$},\label{FHeq}\end{align}
defining the horizontally-averaged enthalpy flux ($F_e$), kinetic energy flux ($F_{KE}$) and viscous flux ($F_{visc}$).
Note that (\ref{Feq}) and (\ref{FHeq}) are equivalent; whether decomposed in the manner of (\ref{Feq}) or the complementary fashion of (\ref{FHeq}), the transport terms must sum to the total luminosity $L$.
$L_{visc}$ represents the total work done by surface forces, whereas $L_{diss}$ represents only the (negative-definite) portion of this that goes into deforming a fluid parcel and hence into heating. 					      

\subsection{Relations between global dissipation rate and convective flux}
For the model described in section \ref{modelsec}, equation (\ref{Phi2}) becomes
\begin{align}\Phi=&-\int_V\bar\rho{s}(\mathbf{u}\cdot\nabla)\bar{T}\,dV\nonumber\\
=&\frac{g}{c_{p,0}}\int_Vs\bar\rho{w}\,dV=\frac{gA}{c_{p,0}}\int_{0}^{d}\frac{F_{conv}}{\bar{T}}\,dz,\label{phiFconv}\end{align}
Often it is assumed that in the bulk of the convection zone, the total heat flux is just equal to the convective flux as defined above (i.e., $F_{conv}\approx{F}$). We show later that this a poor assumption in strongly stratified cases, but it is reasonable for approximately Boussinesq systems. In the case $F_{conv}\approx{F}$, (\ref{phiFconv}) becomes
\begin{equation}
\Phi=\frac{gAF}{c_{p,0}T_0}\int_0^d\frac{1}{1-\beta{z}}\,dz=-L_u\ln(1-\beta{d})
\end{equation}
and 
\begin{equation}\label{lower}
E=-ln(1-\beta{d})=\beta{d}+\ldots\approx\frac{d}{H_{T,0}}.
\end{equation}
However, in strongly stratified cases $F\approx{F_{conv}}+F_{other}$ where $F_{other}=\int_0^{z'}(Q_{buoy}+Q_{diss})\,dz$ from (\ref{Feq}), or alternatively, $F_{other}=F_{p}+F_{KE}+F_{visc}$ from (\ref{FHeq}) (the conductive flux is small in the bulk convection zone). Here $F_{p}=\frac{1}{A}\int_{S_{z'}}wp\,dS$ is the difference between the enthalpy flux $F_e$ and the convective flux $F_{conv}$.  Physically, $F_{other}$ is equivalent to the steady-state transport associated with processes other than the convective flux as defined above. In this case, (\ref{phiFconv}) becomes
\begin{equation}\label{phiFother}
\Phi=\frac{gAF}{c_{p,0}}\int_0^d(1-\frac{F_{other}}{F})\frac{1}{\bar{T}}\,dz,
\end{equation}
where we note that in general $F_{other}$ is a function of depth and $(1-\frac{F_{other}}{F})\geq1$.
A complete theory of convection would specify $F_{other}$ a priori, and thereby constrain the dissipative heating everywhere.
In the absence of such a theory,  we turn to numerical simulations to determine the magnitude of $\Phi$ for strong stratifications.

%%%%%%%

\subsection{Dissipation in simulations: determined by stratification}\label{res1}
\begin{figure}
\includegraphics[scale=1.03]{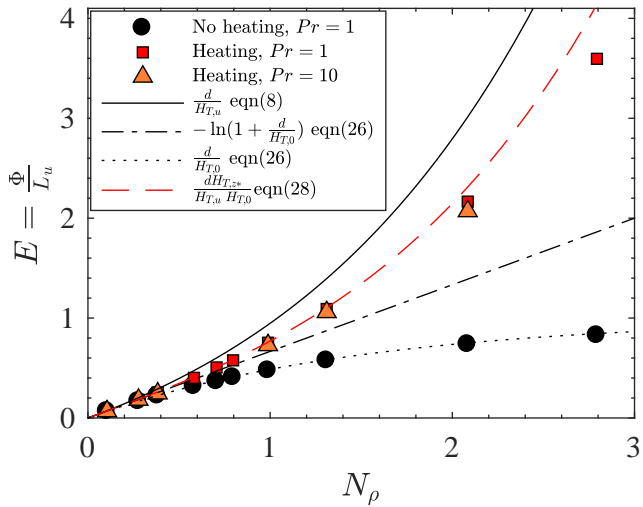}
\caption{$E$ (global dissipative heating rate normalised by the luminosity) against $N_{\rho}$ for $Pr=1$ (red squares) and $Pr=10$ (orange triangles). Cases in which the dissipative heating terms were not included in equation (\ref{energyeq}) are denoted by black circles. The dash-dot line shows the expression given by (\ref{lower}) and the dotted line shows the leading order term of this expression. The solid black line shows the upper bound given by (\ref{bound}) and the dashed red line shows the expression given by (\ref{myE}). The cases with heating agree well with the dashed red line and the cases without heating agree with the dotted black line.}\label{fig1}
\end{figure}

We examine the steady-state magnitude of $\Phi$ for different values of $N_{\rho}$ and $Ra$. 
Figure \ref{fig1} shows the ratio of the global dissipation rate to the luminosity through the layer, $E=\frac{\Phi}{L_u}$, for varying stratifications. First, we highlight the difference between simulations in which the dissipative heating terms were included (red squares) and those where they were not (black circles). At weak stratification, %(low $N_{\rho}$)
there is not much difference in the dissipative heating rate between these cases, but differences become apparent as $N_{\rho}$ is increased. Including the heating terms in a self-consistent calculation leads to a much larger value of $E$ than if $\Phi$ is only calculated after the simulation has run (i.e., if heating is not allowed to feedback on the system). When heating terms are included, the global dissipative heating rate exceeds the flux passing through the system (i.e., $E>1$) when $N_{\rho}>1.22$. 

As expected, the expression for $E$, in the Boussinesq limit, given by (\ref{lower}), is a good approximation to $E$ for small $N_{\rho}$, but vastly underestimates $E$ at large $N_{\rho}$ (see Figure \ref{fig1}, dash-dot line).
In the cases where the heating terms are not included, $E$ cannot exceed unity for all $N_{\rho}$.
This might have been expected, since in this case none of the dissipated heat is returned to the internal energy of the system; instead, the dissipated energy is simply lost (i.e., energy is not conserved). This has the practical consequence that the flux emerging from the top of the layer is less than that input at the bottom.
In these cases $E$ is very well described by the dashed line which is given by $\frac{d}{H_{{T},0}}$, the leading order term from the expression for $E$ in (\ref{lower}).

The theoretical upper bound derived by HMW is shown on Figure \ref{fig1} by the solid black line. It is clear that all of our cases fit well within this upper bound, even at strong stratifications. This upper bound is equivalent to $\frac{d}{H_{{T},u}}$ in this system, where $H_{{T},u}$ is the value of $H_{T}$ on the upper boundary.

Cases in which the heating terms were included are well described by 
\begin{equation}\label{myE}
E=\frac{d}{\tilde{H_T}},
\end{equation}
where
\begin{equation}\label{htdef}
\tilde{H_T} = \frac{H_{T,0}H_{T,u}}{H_{T,z^*}}
\end{equation}
is a modified thermal scale height involving $H_T$ at the top, bottom and at a height $z^*$, defined such that half the fluid (by mass) lies below $z^*$ and half sits above; for a uniform density fluid, $z^*=\frac{d}{2}$. This expression resembles that originally proposed by HMW, on heuristic grounds, for a gas ($E\approx\frac{d}{H_T}$); in our case $H_T$ is not constant across the layer and we find that the combination $\tilde{H_T}$ is the appropriate ``scale height" instead. Like HMW's suggestion, it depends only on the layer depth and temperature scale heights of the system.

For 2D convection, at $Pr=1$ and the $Ra$ considered here, the solutions are steady (time-independent) \citep{VincentYuen1999}; the convection takes the form of a single stationary cell occupying the layer. To assess if the same behaviour occurs for chaotic (time-dependent) solutions, we have included some cases at $Pr=10$ (orange triangles), since then the flow is unsteady. In the cases included here, this unsteady flow is characterised by the breakup of the single coherent convection cell (seen at $Pr=1$); these time-dependent solutions seem also to be well described by the line given by (\ref{myE}). This behaviour is sampled in Figure \ref{figA1}, Supplementary Material, which shows the velocity and entropy fields in a simulation with $Pr=10$, $N_{\rho}=1.31$, $Ra=4.13\times10^8$ and $\hat F_0=0.14$.
At higher $Ra$, the solutions transition to turbulence \citep[see visualisations in e.g.,][]{Rogersetal2003}. 

\subsection{Dissipation in simulations: independent of diffusivities}\label{2p4}
\begin{figure}
\includegraphics[scale=1.03]{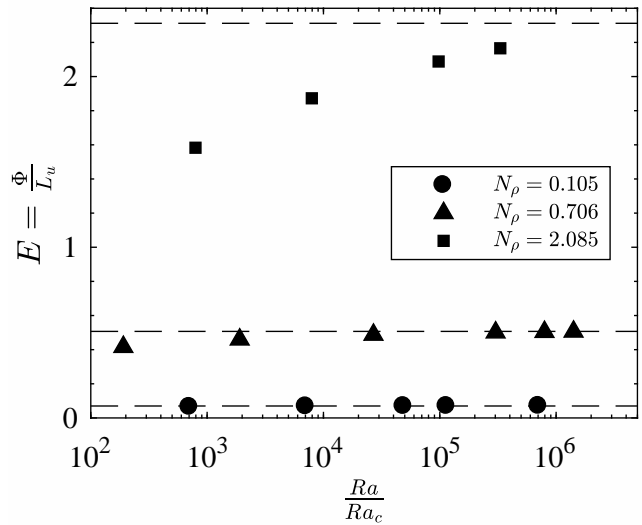}
\caption{$E$ as a function of $\frac{Ra}{Ra_c}$ (where $Ra_c$ is the value of $Ra$ at which convection onsets) for $N_{\rho}=0.105$ (circles), $N_{\rho}=0.706$ (triangles) and $N_{\rho}=2.085$ (squares). In each case, for large enough $Ra$ the value of $E$ asymptotes to the value given by (\ref{myE}), indicated for each $N_{\rho}$ by the horizontal lines. The level of stratification (given by $N_{\rho})$, rather then the diffusion, determines the magnitude of the dissipative heating rate compared to the flux through the layer.}\label{fig2}
\end{figure}

The results of section \ref{res1}, specifically equation (\ref{myE}), suggest that the amount of dissipative heating is determined by the stratification, not by other parameters such as $Ra$. To probe this further, we consider how/if $E$ changes as $Ra$ is varied. Figure \ref{fig2} shows the results for three different stratifications. For $N_{\rho}\approx0.1$, the fluid is close to being Boussinesq and it is clear that $E$ remains constant (and equal to the value given by (\ref{myE})) for many decades increase in $Ra$. This result complements that of HMW obtained from Boussinesq simulations at infinite $Pr$. For increasing $N_{\rho}$, we find that for large enough $Ra$, $E$ approaches the constant given by (\ref{myE}). That $E$ becomes independent of $Ra$ at large enough $Ra$ for all $N_{\rho}$ was also found by \citet{JarvisMcKenzie1980}, albeit for liquids at infinite $Pr$. 

Figure \ref{fig2} indicates that the solutions have to be sufficiently supercritical in order for the theory to be valid. It also suggests that stronger stratifications require simulations to be more supercritical in order to reach the asymptotic regime. (All the simulations displayed in Figure \ref{fig1} approach this asymptotic regime, \emph{except} possibly the uppermost point at $N_{\rho}=2.8$. That simulation has $Ra/Ra_c \approx 9 \times10^{5}$, but it is likely that still higher $Ra$ would yield somewhat greater values of $E$ at this stratification.)

\section{Discussion and conclusion}\label{discussion}
We have demonstrated explicitly that the amount of dissipative heating in a convective gaseous layer can, for strong stratifications, equal or exceed the luminosity through the layer. 
A principal conclusion is that the ratio of the global viscous heating rate to the emergent luminosity is approximated by a theoretical expression dependent only on the depth of the layer and its thermal scale heights.  This ratio, akin to one originally derived for a simpler system by HMW, is given (for the cases studied here) by (\ref{myE}). Interestingly, this relation does not depend on other parameters such as the Rayleigh number. Our simulations confirm that this expression holds for 2D convection in an anelastic gas, provided the convection is sufficiently supercritical.  This regime is attainable in our 2D simulations, and is surely reached in real astrophysical objects, but may be more challenging to obtain in (for example) 3D global calculations \citep[e.g.,][]{FeatherstoneHindman2016,Aubertetal2017}.

\begin{figure}
\includegraphics[scale=1,trim = {0mm 0mm 0mm 0mm}, clip]{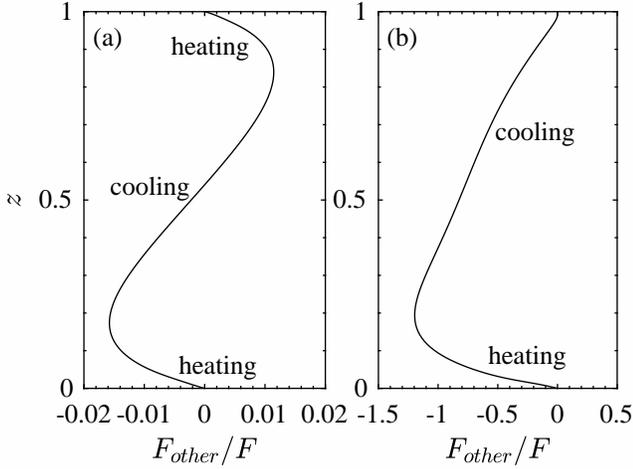}
\caption{Local heating and cooling. $F_{other}$ as a fraction of the total flux through the layer as a function of layer depth for $N_{\rho}=0.1$ in (a) and $N_{\rho}=2.08$ in (b). In (a) the local heating and cooling is only a few percent of the total flux whereas in (b) the local heating and cooling is comparable to the flux through the layer in some parts.}\label{fig3}
\end{figure}

The dissipative heating appears in the local internal energy (or entropy) equation, in the same way as heating by fusion or radioactive decay.  Where it is large, we therefore expect it will modify the thermal structure, just as including a new source of heating or cooling would have done.  It must be reiterated, though, that in a global sense this heating is balanced by equivalent cooling terms; i.e., $L_{diss}$ and $L_{buoy}$ in equation (\ref{Feq}) cancel in a global sense; no additional flux emerges from the upper boundary. Stars are not brighter because of viscous dissipation. Locally, however, these terms do \emph{not} necessarily cancel, as explored in Figure \ref{fig3}.  There we show the net heating and cooling at each depth in two simulations; in Figure \ref{fig3}$a$, the fluid is weakly stratified, and in (b) is has a stratification given by $N_{\rho}=2.08$. In both cases the sum of the terms must be zero at the top and bottom of the layer, but not in between.
Furthermore, in (a) the terms are small compared to the flux through the layer (typically a few \%) but in the strongly stratified case, the local heating and cooling become comparable to the overall luminosity. 
In general, stronger stratifications lead to stronger local heating and cooling in the fluid.

\begin{figure*}
\includegraphics[scale=1,trim = {0mm 0mm 0mm 0mm}, clip ]{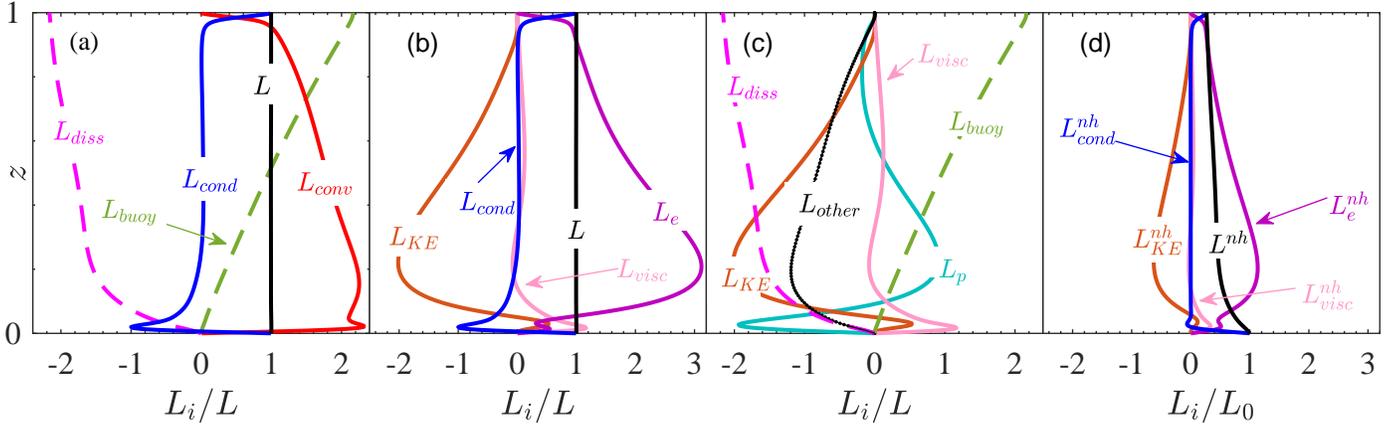}
\caption{(a) Luminosities $L_i$ defined in (\ref{Feq}) and their sum normalised by the total luminosity $L$. (b) Luminosities $L_i$ defined in (\ref{FHeq}) and their sum normalised by the total luminosity $L$. (c) The constituents of $L_{other}=AF_{other}=A\int_0^{z'}(Q_{buoy}+Q_{diss})\,dz=A(F_p+F_{KE}+F_{visc})$. (d) Luminosities $L_i$ defined in (\ref{FHeq}) and their sum normalised by the total luminosity at the bottom boundary $L_0$ in the case where heating terms are not included. The luminosities in (d) are significantly smaller than the equivalent ones when heating terms were included (see (b)).}\label{fig4}
\end{figure*}

In a steady state the imbalance between this local heating and cooling is equivalent to certain transport terms as discussed in section \ref{modelsec}; these are assessed for our simulations in figure \ref{fig4} where the terms are plotted as luminosities and labelled correspondingly. Turning first to Figure \ref{fig4}$a$, we show the components of the total flux of thermal energy (as described by (\ref{Feq})), namely $L_{conv}$, $L_{cond}$, $L_{buoy}$ and $L_{diss}$. The conductive flux is small throughout the domain except in thin boundary layers and the dissipative heating ($L_{diss}$) is comparable to the convective flux ($L_{conv}$) throughout the domain. The sum of the four transport terms is shown as the black line ($L$) and is constant across the layer depth, indicating thermal balance.
Figure \ref{fig4}$b$ assesses the total energy transport using the complementary analysis of (\ref{FHeq}), using $L_{KE}$, $L_{cond}$, $L_e$ and $L_{visc}$. The primary balance is between the positive $L_e$ and the negative $L_{KE}$. Viewed in this way, the viscous flux ($L_{visc}$) is small except near the lower boundary, but (as discussed in section \ref{modelsec}) this does not necessarily mean the effect of viscous dissipation is also small.
In figure \ref{fig4}$c$ we highlight the equivalence of some transport terms, by showing the term $AF_{other}$ together with its different constituent terms from either the total or thermal energy equations. As expected, $AF_{other}$ is the same in both cases; it is the sum of $L_{diss}$ and $L_{buoy}$, or equivalently, it is the sum of $L_{p}$, $L_{KE}$ and $L_{visc}$. That is, changes in the dissipative heating are reflected not just in $Q_{diss}$ (if analysing internal energy) or $F_{visc}$ (if analysing total energy); the other transport terms ($F_{KE}$, $F_p$, $F_e$, $F_{conv}$, $Q_{buoy}$) also change in response.
To emphasise the importance of dissipative heating in modifying the transport terms, we include in Figure \ref{fig4}$d$, $L_{KE}^{nh}$ , $L_{e}^{nh}$ , $L_{cond}^{nh}$  and $L_{visc}^{nh}$ i.e., the kinetic energy, enthalpy, conductive and viscous fluxes (expressed as luminosities) respectively, in the case where heating terms were not included. It is clear that these are much smaller than in the equivalent simulation with heating (Figure \ref{fig4}$b$), demonstrating explicitly that the inclusion of dissipative heating influences the other transport terms.
In particular, the maximum value of the kinetic energy flux is 3.2 times larger when the heating terms are included. The black line in Figure \ref{fig4}$d$ shows that when heating is not included the flux emerging at the upper boundary is smaller than the flux imposed at the lower boundary; in this case it is approximately $27\%$ of $L$.

The local heating and cooling (or, equivalently, the transport term $F_{other}$ that must arise from this in a steady state) described above is not included in standard 1D stellar evolution models, and we do not yet know what effects (if any) would arise from its inclusion. 
In some contexts they may be negligible; the total internal energy of a star is enormously greater than its luminosity $L\star$, so even internal heating that exceeds $L\star$ may not have a noticeable effect on the gross structure. If, however, this heating is concentrated in certain regions (e.g., because of spatially varying conductivity) or occurs in places with lower heat capacity, its impact may be more significant.
If the results explored here also apply to the full 3D problem with rotation and magnetism -- which clearly must be checked by future calculation -- then the total dissipative heating is determined non-locally, dependent as it is on the total layer depth.  Simple modifications to the mixing-length theory (which is determined locally) may not then suffice to capture it. We have begun to explore these issues by modification of a suitable 1D stellar evolution code, and will report on this in future work.

\acknowledgments
We acknowledge support from the European Research Council under ERC grant agreements No. 337705 (CHASM). The simulations here were carried out on the University of Exeter supercomputer, a DiRAC Facility jointly funded by STFC, the Large Facilities Capital Fund of BIS and the University of Exeter. We also acknowledge PRACE for awarding us access to computational resources Mare Nostrum based in Spain at the Barcelona Supercomputing Center, and Fermi and Marconi based in Italy at Cineca. We thank the referee for a thoughtful review that helped to improve the manuscript.

\appendix
\section{Simulation parameters}
\begin{deluxetable}{CCCCCC}
\tablecaption{Simulation parameters used in figures \ref{fig1}-\ref{fig4} \label{table1}}
\tablecolumns{6}
\tablenum{1}
\tablewidth{0pt}
\tablehead{
\colhead{$Pr$} &
\colhead{$N_{\rho}$} & \colhead{$Ra$} & \colhead{$\hat{F_0}$} &
\colhead{$E$} & \colhead{Figure}
}
\startdata
1 & 0.1050 & 3.83 \times 10^5 & 3.26 \times 10^{-4} & 0.0630 & 2\\
1 & 0.1050 & 3.83 \times 10^6 & 3.26 \times 10^{-3} & 0.0662 & 2\\
1 & 0.1050 & 2.63 \times 10^7 & 2.24 \times 10^{-2} & 0.0678 & 2\\
1 & 0.1050 & 6.13 \times 10^7 & 5.22 \times 10^{-2} & 0.0682 & 2\\
1 & 0.1050 & 3.83 \times 10^8 & 3.26 \times 10^{-1} & 0.0689 &1-3\\
1 & 0.2776 & 6.58 \times 10^7 & 5.60 \times 10^{-2} & 0.1828 &1\\
1 & 0.3828 & 8.77 \times 10^7 & 7.47 \times 10^{-2} & 0.2557 &1\\
1 & 0.5819 & 8.01 \times 10^7 & 1.07 \times 10^{-3} & 0.4014 &1\\
1 & 0.7060 & 6.65 \times 10^4 & 3.62 \times 10^{-3} & 0.4159 &2\\
1 & 0.7060 & 6.65 \times 10^5 & 3.62 \times 10^{-2} & 0.4594 &2\\
1 & 0.7060 & 9.36 \times 10^6 & 1.24 \times 10^{-4} & 0.4875 &2\\
1 & 0.7060 & 1.05 \times 10^8 & 2.64 \times 10^{-3} & 0.5008 &2\\
1 & 0.7060 & 2.72 \times 10^8 & 3.62 \times 10^{-3} & 0.5038 &2\\
1 & 0.7060 & 4.88 \times 10^8 & 4.40 \times 10^{-3} & 0.5057 &1-2\\
1 & 0.7967 & 1.03 \times 10^8 & 1.37 \times 10^{-3} & 0.5770 &1\\
1 & 0.9887 & 1.20 \times 10^8 & 1.60 \times 10^{-3} & 0.7533 &1\\
1 & 1.3104 & 8.45 \times 10^7 & 1.12 \times 10^{-3} & 1.0908 &1\\
1 & 2.0846 & 1.33 \times 10^5 & 7.24 \times 10^{-3} & 1.5830 &2\\
1 & 2.0846 & 1.33 \times 10^6 & 2.68 \times 10^{-3} & 1.8726 &2\\
1 & 2.0846 & 1.63 \times 10^7 & 2.17 \times 10^{-4} & 2.0882 &2\\
1 & 2.0846 & 5.44 \times 10^7 & 7.24 \times 10^{-4} & 2.1656 &1-4\\
1 & 2.7938 & 1.23 \times 10^8 & 1.63 \times 10^{-3} & 3.5951 &1\\     
10 & 0.1050 & 2.63 \times 10^7 & 1.43 \times 10^{-1} & 0.0668&1\\
10 & 0.2776 & 1.15 \times 10^8 & 9.78 \times 10^{-3} & 0.1822&1\\
10 & 0.3828 & 3.25 \times 10^6 & 1.59 \times 10^{-1} & 0.2454&1\\
10 & 0.9887 & 1.37 \times 10^9 & 4.66 \times 10^{-1} & 0.7413&1\\
10 & 1.3104 & 4.13 \times 10^8 & 1.40 \times 10^{-1} & 1.0594&1\\
\enddata
\end{deluxetable}

\begin{figure*}
\includegraphics[scale=1,trim = {0mm 0mm 0mm 0mm}, clip ]{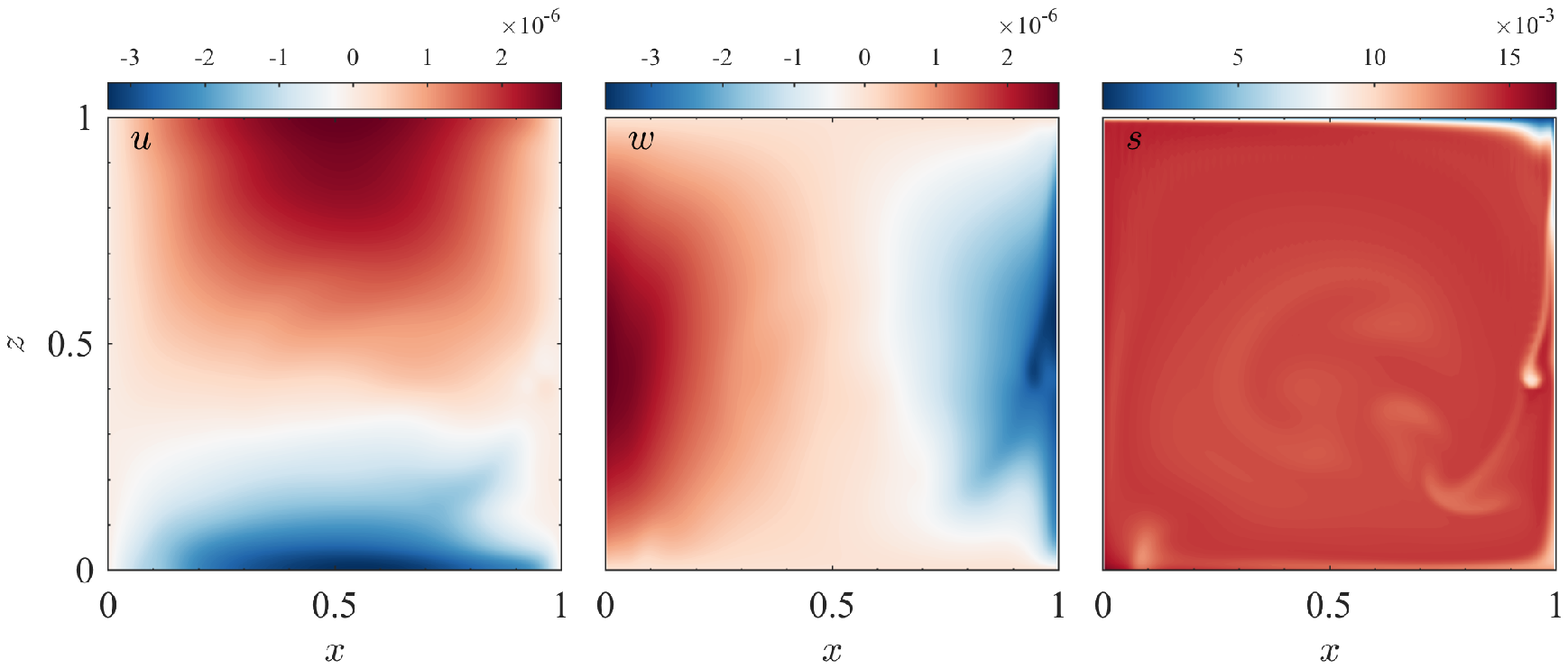}
\caption{Snapshot of the statistically-steady velocity components, $u$ and $w$, and the entropy field $s$ for a simulation with $Pr=10$, $N_{\rho}=1.31$, $Ra=4.13\times10^8$ and $\hat F_0=0.14$. The values of $u$ and $w$ are nondimensionalized using the box height as a typical length scale and $\tilde t=\frac{\nu}{gd}$ as a characteristic time scale. $s$ is given in units of $c_p$.}\label{figA1}
\end{figure*}

\end{document}